\documentclass{ceab}
\usepackage{epsfig}
\usepackage{graphicx}
\usepackage{ceabbib}
\usepackage{txfonts}

\begin{document}
\title{Temporal variations of Coriolis-turns in the photosphere}
\author{J. MURAK\"OZY and A. LUDM\'ANY
\vspace{2mm}\\
\it Heliophysical Observatory of the Hungarian Academy of Sciences,\\
\it H-4010 Debrecen P.O.Box 30. Hungary\\}

\maketitle
 
\begin{abstract}
Correlations have been studied between longitudinal and latitudinal shifts of sunspot groups in the period 1986-1998. This study is based on the Debrecen Photoheliographic Data (DPD). The latitudinal distribution of the correlation values exhibits the expected Coriolis-pattern on the entire period: negative/positive correlations in the northern/southern hemispheres, gradually disappearing toward the equator. The yearly distributions, however, show an interesting cycle dependence: after the unambiguous Coriolis patterns during the growing phase of the cycle the latitudinal dependence of the correlations gets ambiguous at the activity maximum. This may be related to the Gnevyshev gap, the depression of the cycle time-profile at the maximum.
\end{abstract}

\def\gore{Varying Coriolis-motions}
\keywords{sunspot motions, cyclic variations}

\section{Introduction}
The Coriolis effect in the solar photosphere can be detected by computing correlation between the longitudinal and latitudinal motions of photospheric features. This has been done first by Ward (1965) who computed the $\langle v_{\theta} v_{\phi} \rangle$ covariance values from the Greenwich sunspot data and interpreted the results as signatures of the $\langle v_{i} v_{j} \rangle$ turbulent velocity covariances called Reynolds stresses. His results were corroborated in several later studies by Coffey and Gilman (1969), Gilman and Howard (1984), Howard (1991), Pulkkinen and Tuominen (1998). The latitudinal variation of the correlation was always found to be nearly linear. The effect can also be pointed out in higher atmospheric layers, in chromospheric bright mottles (Schr\"oter and W\"ohl, 1976; Belvedere et al., 1976) and in the low corona (Vr\v{s}nak et al. 2003). 

The present work studies the variations of latitudinal correlation dependence. Some expectations may be formulated. The Coriolis effect itself cannot depend on time but it may be weak or absent in minimum activity at the equator, on the other hand, it may be distorted by some varying circumstances during the cycle.

\section{Analysis of data}

The work is based on the Debrecen Photoheliographic Data (DPD, Gy\H ori et al., 2007), a detailed sunspot catalogue. At the time of the analysis it covered 13 years between 1986 and 1998, the entire solar cycle 22. It provides the position and area data of all sunspots and sunspot groups on a daily basis. Correlation coefficients have been computed for all sunspot groups in the following way. In order to avoid the false shifts on account of the emerging new spots the maximum phases were selected for all sunspot groups. Only those groups were considered which had their maximum area on the visible disc and for not more than three days prior to and following the maximum phase. $\Delta L_{i}$ longitudinal and $\Delta B_{i}$ latitudinal diurnal shifts were determined for these periods in such a way that the differences between the daily data have been normalized to values of 24 hours (because the observations are taken in different moments of the days) and the correlation coefficient was computed between these two data serii for the period of observability within $\pm 70^{o}$ from the central meridian. Differential rotation has not been considered. All selected sunspot groups received a correlation coefficient and these were averaged for $5^{o}$ latitudinal bins. The result, the latitudinal dependence of ($\Delta L_{i}$,$\Delta B_{i}$) correlations is depicted in Figure 1. It is similar to that published by Latushko (1993), who obtained similar dependence for magnetograms. The distribution published by Pulkkinen and Tuominen (1998) is basically the same but this paper deals with covariances. The correlation disappears at the equator, as is expected, and its absolute value grows monotonously toward higher latitudes, it is negative/positive in the northern/southern hemispheres, i.e. a longitudinal forward motion is probably connected with the equatorward latitudinal motion. 

\begin{figure}
\begin{center}
\epsfig{file=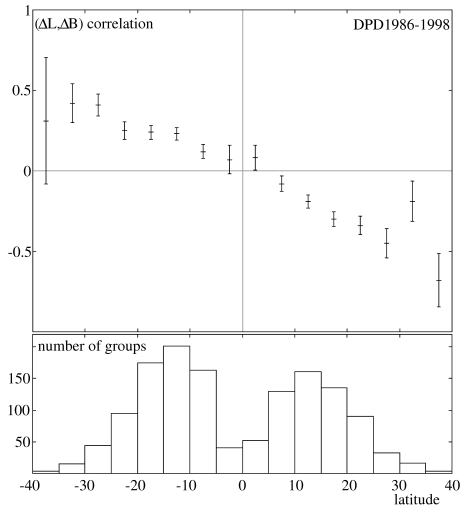,width=6cm}%
\end{center}
\caption{Upper panel: latitudinal distribution of sunspot motion correlation for the years 1986-1998, lower panel: numbers of sunspot groups within the $5^{o}$ latitudinal stripes.}
\end{figure} 

The aim of the present work is to reveal any temporal variation in this pattern. Figure 2 shows annual distributions separately around the cycle maximum. Years of low activity are disregarded here because close to the equator no characteristic patterns are obtained similarly to the results of Gilman and Howard  (1984) and Balthasar et al. (1986). The distribution exhibits unexpected variations around the maximum, the longitudinal dependence is unanimous in the year 1988 but it weakens in the year 1989 and strengthens in 1990. This cannot be the result of smaller statistics because the number of sunspots is higher than in 1988 and only slightly smaller than in 1989, moreover the activity belt occupied by sunspots is the widest around the maximum, so we would expect the opposite case: the most unambiguous monotonic decreasing trend in 1989-91. Its lack in 1990 may indicate some deeper phenomenon.

The shape of the cycle 22 can be followed in Figure 3. It is remarkable that the weakening of the latitude dependence coincides with the local depression of the cycle profile at the maximum, the so called Gnevyshev gap. This property has been first described by Gnevyshev (1967) and analysed by several authors (Bazilevskaya et al., 2000; Storini et al., 2003). To check this impression, the correlations of the fitted lines with the distributions in the frames of Figure 3 have also been computed and these values are indicated with crosses in Figure 3. The similarity of the activity curve and these correlation values is surprising: both profiles have a drop in 1990, at the Gnevyshev gap. As a further possibly relevant piece of information, the moments of the magnetic polarity reversals are alo indicated by using the data of Makarov and Makarova (1996).

\begin{figure}
\begin{center}
\epsfig{file=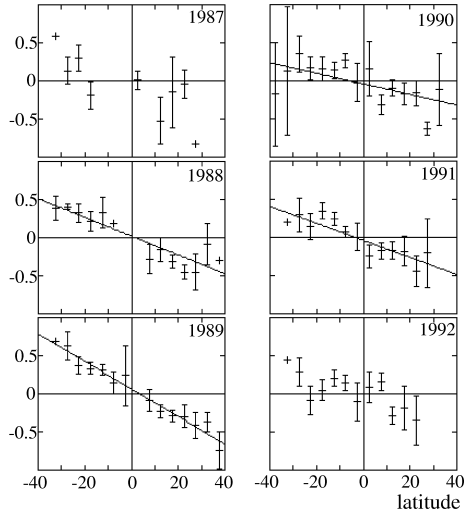,width=6cm}
\end{center}
\caption{Latitudinal distributions of the $(\Delta L, \Delta B)$ correlations on a yearly basis in 1987-1992.}
\end{figure}

\begin{figure}
\begin{center}
\epsfig{file=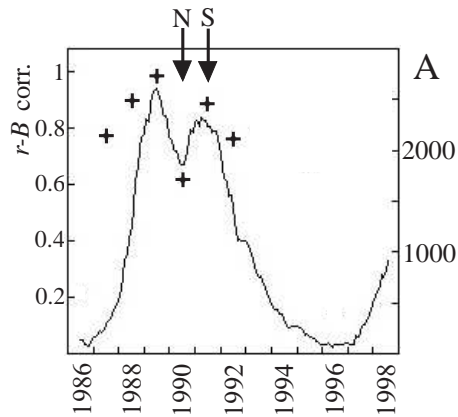,width=6cm}
\end{center}
\caption{Shape of cycle 22 by smoothed monthly mean sunspot area data (from DPD), {\it A} denotes the area in MSH (Millionths of the Solar Hemisphere). The crosses show the correlations between the $(\Delta L, \Delta B)$ correlation values here denoted by {\it r}   and the {\it B} heliographic latitude in each year. The times of magnetic polarity reversals at the northern and southern poles are indicated with arrows.}
\end{figure} 

\section{Discussion, variations around maxima}

The earlier interpretation of the obtained patterns was given by referring to the Reynolds stresses. Most of the earlier attempts used covariances between the latitudinal and longitudinal shifts and it was assumed that the results were signatures of these turbulent features which were held responsible for the equatorward angular momentum transport. However, D'Silva and Howard (1995) pointed out that this interpretation is not the only one possible, the magnetic field ropes can be subjects of Coriolis turns without any dragging by the ambient plasma. Thus the present work doesn't consider the magnitude of the effect (which is interesting to esteem the angular momentum transport) but instead, it focuses on the temporal variations which may be related to the evolution of the magnetic fields during the cycle.

The distribution averaged on long intervals depicted in Figure 1 exhibits the pattern which can be expected on account of the Coriolis effect. As for the yearly distributions, the expectations are based on the Sp\"orer's law: when active regions are at higher latitudes, the Coriolis effect should be more expressed than at the end of the cycle decay, close to the equator. In fact, this is the case, so the minimum years are not presented here as they are not informative. One may expect, however, that in the years of activity maximum, when the active regions are spread over the widest latitudinal belts, the distribution of shift-correlations is the most comparable to the distribution averaged over the entire cycle (Figure 1) but this is not the case. In certain years at maximum the regular Coriolis pattern may significantly weaken. 

Variations of the differential rotation cannot play a role in the variations of the $(\Delta L, \Delta B)$ correlation. The individual correlation values are computed for individual sunspot groups, for a limited period of their observability and by using diurnal shifts. During these short periods of visibility the differential rotation is constant and the diurnal shifts are completely independent from the actual rotational velocity values.

The fluctuation of the two time profiles of Figure 3 at maxium may somehow be related but no causal order can be presumed as yet, moreover, the coincidence of the Gnevyshev gap and the weakened Coriolis-pattern may happen even by chance.

The role of the polarity reversal of the main magnetic dipole field can apparently be excluded. As is perceivable in Figure 3, the polarity reversals (at both poles) follow rather than preceed the local depression at the maximum, thus no causal connection can be conceived between these fluctuations and the turnover of the global dipole.

A further interpretation may be based on the idea reported by Bazilevskaya et al. (2000). They assumed that the Gnevyshev gap may be a result of a quasi-biennial fluctuation superposed onto the cycle. These fluctuations, also called mid-term fluctuations, may be found in a wide variety of phenomena covering the region from the tachoclyne zone through the solar surface and interplanetary space up to the terrestrial environment and cosmic rays but probably most of them are independent from each other. Their periods are also different and varying (Mursula et al., 2003; Ivanov et al., 2002). In the present case some subsurface fluctuations may be examined, primarily the shear oscillation at the tachoclyne zone (Howe and Christensen-Dalsgaard, 2000). If this fluctuation can interact with the deep turbulent motion field, e.g. the giant convection cells, then this interaction may modify the motions of magnetic field ropes as well. The Coriolis effect itself can only depend on the latitude but not on time. 
If it does vary in time then it may be the signature of a varying process which presents varying conditions for the motions of magnetic fields. If the latitudinal pattern of the Coriolis effect varies then a subsurface redistribution process may be conjectured. This may be the target of further studies.

\section*{Acknowledgements}
This work was supported by the ESA PECS project No. 98017. We express our appreciation to L. Gy\H ori and T. Baranyi for their work in producing the DPD catalogue.

\section*{References}
\begin{itemize}
\small
\itemsep -2pt
\itemindent -20pt
\item[] Balthasar, H.; V\'azquez, M.; W\"ohl, H., 1986, \aap, 155, 87
\item[] Bazilevskaya, G. A., Krainev, M. B., Makhmutov, V. S., Fl\"uckiger, E. O., Sladkova, A. I. \& Storini, M., 2000, \solphys, 197, 157 
\item[] Belvedere, G., Godoli, G., Motta, S., Patern\`o, L. \& Zappal\`a, R. A., 1976, \solphys, 46, 23
\item[] Coffey, H. E. \& Gilman, P. A., 1969, \solphys, 9, 423 
\item[] D'Silva, S.\& Howard, R. F., 1995, \solphys, 159, 63 
\item[] Gilman, P. A. \& Howard, R., 1984, \solphys, 93, 171 
\item[] Gnevyshev, M. N., 1967, \solphys, 1, 107 
\item[] Gy\H ori, L., Baranyi, T., Ludm\'any, A. et al., 2007, Debrecen Photoheliographic Data for 1986-1998, see: http://fenyi.solarobs.unideb.hu/DPD/index.html
\item[] Howard, R. F., 1991, \solphys, 131 259
\item[] Howe R.; Christensen-Dalsgaard J., 2000, Science 287, 2456
\item[] Ivanov, E. V.; Obridko, V. N.; Shelting, B. D.,2002, ESA SP-506, Vol. 2., 847
\item[] Latushko, S., 1993, \solphys, 146, 401  
\item[] Makarov, V.I.; Makarova, V.V., 1996, \solphys, 163, 267
\item[] Mursula, K. Zieger, B., and Vilppola, J. H., 2003, \solphys, 212, 201
\item[] Pulkkinen, P. \& Tuominen, I., 1998, \aap, 332, 755
\item[] Schr\"oter, H. \& W\"ohl, H., 1976, \solphys, 49, 19 
\item[] Storini, M., Bazilevskaya, G. A., Fluckiger, E. O., Krainev, M. B., Makhmutov, V. S. \& Sladkova, A. I., 2003, Adv. Space Res., 31, 4, 895
\item[] Vr\v{s}nak, B.; Braj\v{s}a, R.; W\"ohl, H.; Ru\v{z}djak, V.; Clette, F.; Hochedez, J.-F., 2003, \aap, 404, 1117 
\item[] Ward F., 1965, \apj, 141, 534
\end{itemize}

\end{document}